\def\CO2{CO$_{2}$} 
\def\P42{$P4_{2}/mnm$}

\def\I42{$I\bar{4}2d$} 

\documentclass[%
reprint,
superscriptaddress,
amsmath,amssymb,
prl,
]{revtex4-1}

\usepackage{xcolor}
\usepackage{graphicx}
\usepackage{dcolumn}
\usepackage{bm}

\usepackage{graphicx}
\usepackage{subcaption}
\usepackage{multirow}
\usepackage{float}

\begin{document}


\title{{\it Ab initio} determination of the phase diagram of CO$_2$ at high pressures and temperatures} 
\author{Beatriz H. Cogollo-Olivo}
  \affiliation {Universidad de Cartagena, Doctorado en Ciencias F\'isicas \\
  Cartagena de Indias, 130001, Colombia}  
\author{Sananda Biswas} 
   \affiliation{Institut f\"{u}r Theoretische Physik, Goethe-Universit\"{a}t Frankfurt, 60438 Frankfurt am Main, Germany}
\author{Sandro Scandolo}
  \affiliation {The Abdus Salam International Centre for Theoretical Physics (ICTP) \\
  Trieste, Strada Costiera 11, 34151, Italy}
\author{Javier A. Montoya}
  \affiliation {Universidad de Cartagena, Instituto de Matem\'aticas Aplicadas \\
  Cartagena de Indias, 130001, Colombia}
    
\date{\today}

\begin{abstract}
The experimental study of the CO$_2$ phase diagram is hampered by strong kinetic effects leading to wide regions of metastability and to large uncertainties in the location of some phase boundaries. Here we determine CO$_2$'s thermodynamic phase boundaries by means of {\it ab initio} calculations of the Gibbs free energy of several solid phases of CO$_2$ up to 50 Gigapascals. Temperature effects are included in the quasi-harmonic approximation. Contrary to previous suggestions, we find that the boundary between molecular forms and the non-molecular phase V has indeed a positive slope and starts at 21.5 GPa at $T$ = 0 K.  A triple point between phase IV, V, and the liquid phase is found at 35 GPa and 1600 K, indicating a broader region of stability for the non-molecular form, than previously thought.  The experimentally determined boundary line between CO$_{2}$-II and CO$_{2}$-IV phases is reproduced by our calculations, indicating that kinetic effects do not play a major role in that particular transition. Our results also show that CO$_{2}$-III is stabilized at high temperature  and  its  stability  region  coincides  with  the $P-T$ conditions where phase VII has been reported experimentally; instead, phase  II  is  the  most  stable  molecular  phase  at  low  temperatures,  extending  its  region  of stability  to  every $P-T$ condition  where  phase  III is reported experimentally.
\end{abstract}


\maketitle

\begin{figure}[t!]
\centering
\includegraphics[width=1.0\columnwidth]{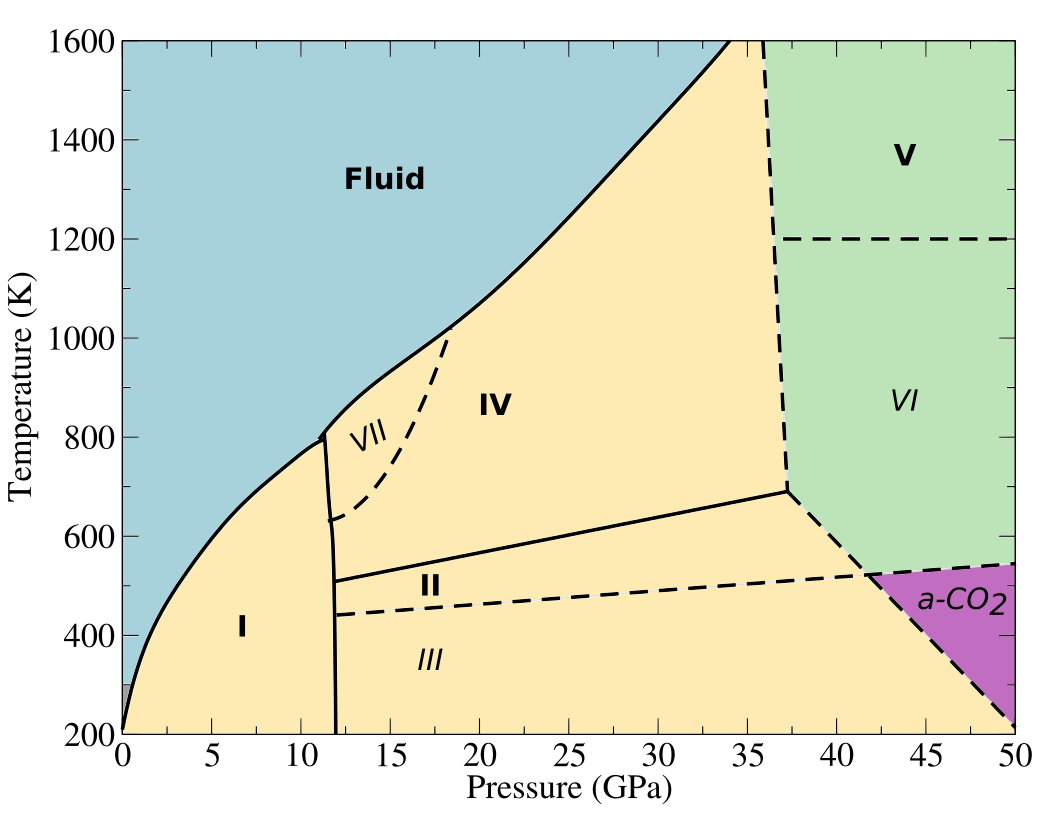}
\caption{High-pressure CO$_{2}$ phase diagram adapted from Ref. \onlinecite{Datchi-2016}. Yellow, green, blue, and purple areas correspond to the molecular, non-molecular, fluid, and amorphous forms of CO$_{2}$, respectively. Solid lines correspond to thermodynamic phase boundaries, while dashed lines are kinetic boundaries. A shaded area on top of the molecular phases spans the experimentally restricted region where the true molecular-to-non-molecular thermodynamic boundary should be present. Names in bold and italic indicate thermodynamic and metastable phases, respectively.}
\label{fig:1}
\end{figure}

Widely studied during the past years, carbon dioxide (CO$_{2}$) is a fascinating system that, despite its simple molecular form at ambient conditions, exhibits a rich polymorphism, with up to seven crystalline  structures reported experimentally in addition to an amorphous form (see Fig. \ref{fig:1}). At room temperature the molecular gas transforms into a liquid at 7.5 MPa which then solidifies at 0.5 GPa into CO$_{2}$-I, a molecular crystal with space group $Pa\bar{3}$ \cite{Simon-1980,Downs-1998}. By further increasing pressure at ambient temperature, CO$_{2}$-I transforms to the orthorhombic phase III ($Cmca$ space group) above 10 GPa, with a minimal volume change \cite{Aoki-1994}. A recent theoretical study has provided insights into the microscopic mechanism of the $Pa\bar{3}$-to-$Cmca$ transition \cite{Gimondi-2017}. Heating compressed CO$_{2}$-III above $\sim$ 470 K \cite{Iota-2001,Gorelli-2004} leads to the transformation into phase II. However this transition is not reversible: CO$_{2}$-II can be recovered at ambient temperature while pressurized, suggesting that CO$_{2}$-III is a kinetic product of the compression of CO$_{2}$-I and not a stable phase \cite{Iota-2001,Santoro-2006}. A recent theoretical study seems to contradict this picture by concluding that CO$_{2}$-III is more stable than CO$_{2}$-II in the low temperature region \cite{Han-2019}, still, that theoretical approach fails to reproduce key experimental observations that show that the transition from phase III to II is observed to happen at temperatures around 475K at 19 GPa, which would be physically impossible if the region of stability of phase III extended up to temperatures around $\sim$ 570 K at that same pressure, as the authors in \cite{Han-2019} claim. Initially described as a structure with carbon in an unconventional six-fold coordination, phase II was interpreted as an intermediate state between the molecular and the extended solid form of CO$_{2}$ \cite{Yoo-2002}, however, subsequent studies disproved the existence of such an intermediate bonding state and identified the structure of phase II as composed of undistorted molecules, with space group $P4_{2}/mnm$ \cite{Datchi-2014}. CO$_{2}$-II transforms into CO$_{2}$-IV when it is heated in the 500 -- 720 K range, depending on pressure \cite{Iota-2001,Santoro-2006}. Phase IV, similarly to phase II, was also initially interpreted as an intermediate bonding state \cite{Iota-2007}, here again, this interpretation was disproved by showing experimentally that CO$_{2}$-IV is still composed of well defined linear molecules and that its crystalline structure is rhombohedral with space group $R\bar{3}c$ \cite{Datchi-2009}. At higher temperatures, an intermediate phase between CO$_{2}$-I and CO$_{2}$-IV was observed by heating CO$_{2}$-I to $\sim$ 950 K and compressing it up to 20 GPa. The crystal structure of this molecular high-temperature stable phase (CO$_{2}$-VII) belongs to space group $Cmca$. \cite{Giordano-2007}. Despite the fact that CO$_{2}$-VII and CO$_{2}$-III have the same space group, some differences in their Raman spectra and in their lattice parameters suggested that their structures might be quantitatively and qualitatively different \cite{Giordano-2007}. However, a recent theoretical study has shown that CO$_{2}$-III and CO$_{2}$-VII belong to the same configurational energy minimum and that CO$_{2}$-III is a low temperature metastable manifestation of CO$_{2}$-VII \cite{Sontising-2017}.

The non-molecular CO$_{2}$-V phase was first synthesized by laser heating CO$_{2}$-III above 40 GPa and 1800 K \cite{Iota-1999}, and its crystalline structure has been now determined as a fully tetrahedral partially collapsed cristobalite-like structure, with space group $I\bar{4}2d$ \cite{Santoro-2012,Datchi-2012}. By compressing CO$_{2}$-II to 50 GPa at 530 and 650 K, another non-molecular form of carbon dioxide (CO$_{2}$-VI) was obtained \cite{Iota-2007}. Its vibrational spectra is consistent with those of metastable layered tetrahedral structures, as shown in \cite{Lee-2009}. In addition to the molecular and non-molecular phases, an amorphous form of carbon dioxide (a-CO$_{2}$) was observed upon compressing CO$_{2}$-III in the pressure range from 40 to 48 GPa at room temperature \cite{Santoro-2006}. The microscopic structure of a-CO$_{2}$ has been explained as a frustrated mixture of 3- and 4-fold coordinated carbon atoms, in an intermediate metastable form towards fully tetrahedral coordination \cite{Montoya-2008}. Finally, CO$_{2}$-V, being the only thermodynamically stable non-molecular form known so far, has been reported to dissociate into elemental carbon (diamond) and oxygen ($\epsilon$-O$_{2}$) at pressures between 30 and 80 GPa, and temperatures above 1700 K, \cite{Tschauner-2001,Litasov-2011}. Nevertheless, there are still conflicting conclusions regarding the CO$_{2}$ dissociation, since more recent theoretical \cite{Teweldeberhan-2013} and experimental \cite{Dziubek-2018} works have not observed a transition from the non-molecular CO$_{2}$-V phase into a dissociated state. Moreover, Dziubek \textit{et al.} \cite{Dziubek-2018} confirmed that CO$_{2}$-V is the only stable phase among all known non-molecular forms of carbon dioxide, as already proposed by a previous experimental work \cite{Seto-2010} as well as by theoretical structural searches in the previous decade. The currently accepted phase diagram including all the mentioned forms of solid CO$_{2}$ along with the region where it becomes a fluid is shown in Fig. \ref{fig:1}. 

In this particular system, very strong kinetic effects hinder the experimental determination of the phase boundaries, while the small size of the samples makes structure determinations quite difficult. As a consequence, several questions remain open regarding the nature of the phase boundaries and the stability of the phases reported in Fig. \ref{fig:1}. In the molecular portion of the phase diagram open questions include the relative stability of CO$_2$-II and CO$_2$-III at low temperature, and the nature of CO$_2$-VII, in particular its structural relationship with CO$_2$-III. At higher pressures one of the fundamental questions is the location of the phase boundary between molecular and non-molecular phases. Santoro \textit{et al.} for example, proposed a phase diagram where the boundary between molecular and non-molecular phases at room temperature is located at 20 GPa, roughly half-way between the lowest pressure of quenching and the pressure of synthesis for this phase \cite{Yong-2016, Santoro-2004}. Moreover, the {\em kinetic} boundary between CO$_2$-III and the a-CO$_{2}$ non-molecular structure, i.e. the $P-T$ region where the transformation occurs upon compression, has a negative slope \cite{Santoro-2009}, while basic thermodynamic considerations suggest that the slope of the true phase boundary should be positive \cite{Santoro-2004}. Theoretical determinations of the molecular/non-molecular boundary at zero temperature, based on {\it ab initio} electronic structure methods, predict transition pressures in the range between 18 and 21 GPa when going from both CO$_{2}$-II and CO$_{2}$-III to the non-molecular forms \cite{Oganov-2008,Gohr-2013,Yong-2016}. 

In this paper, we extend  the theoretical determination of the phase diagram of CO$_2$ to finite temperatures for all stable phases except CO$_2$-I. Phase boundaries between the molecular phases II, III, and IV, and the non-molecular phase V are calculated based on an {\it ab initio} approach; for the determination of free energies the vibrational contributions are treated in the quasi-harmonic approximation. 

\begin{figure}[b!]
\centering
\includegraphics[width=1.0\columnwidth]{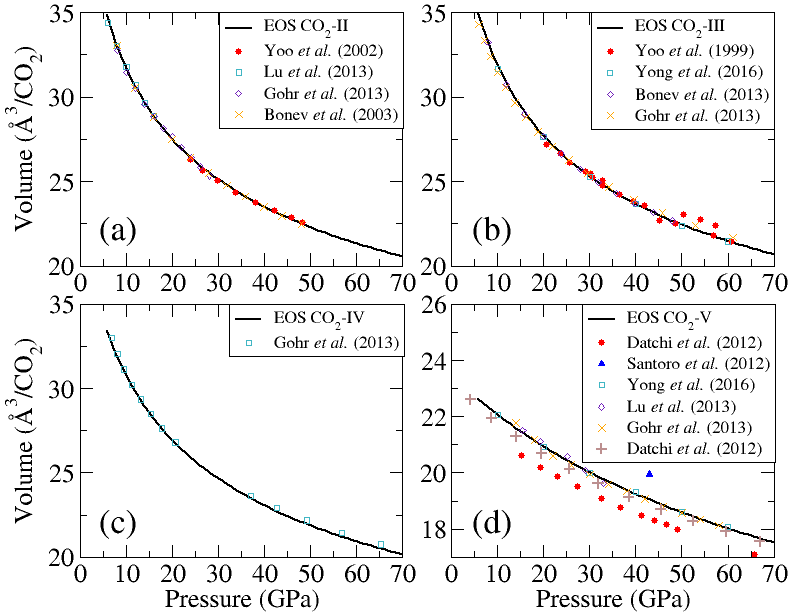}
\caption{Pressure-volume relation of phases (a) II, (b) III, (c) IV, and (d) V of CO$_{2}$ at room temperature are shown in black solid lines. For each case, reported values from experimental (red circles) and theoretical (blue squares, purple diamonds and yellow crosses) studies for the different phases are displayed as well.}
\label{fig:2}
\end{figure}

\begin{figure*}
\centering
\includegraphics[width=2.0\columnwidth]{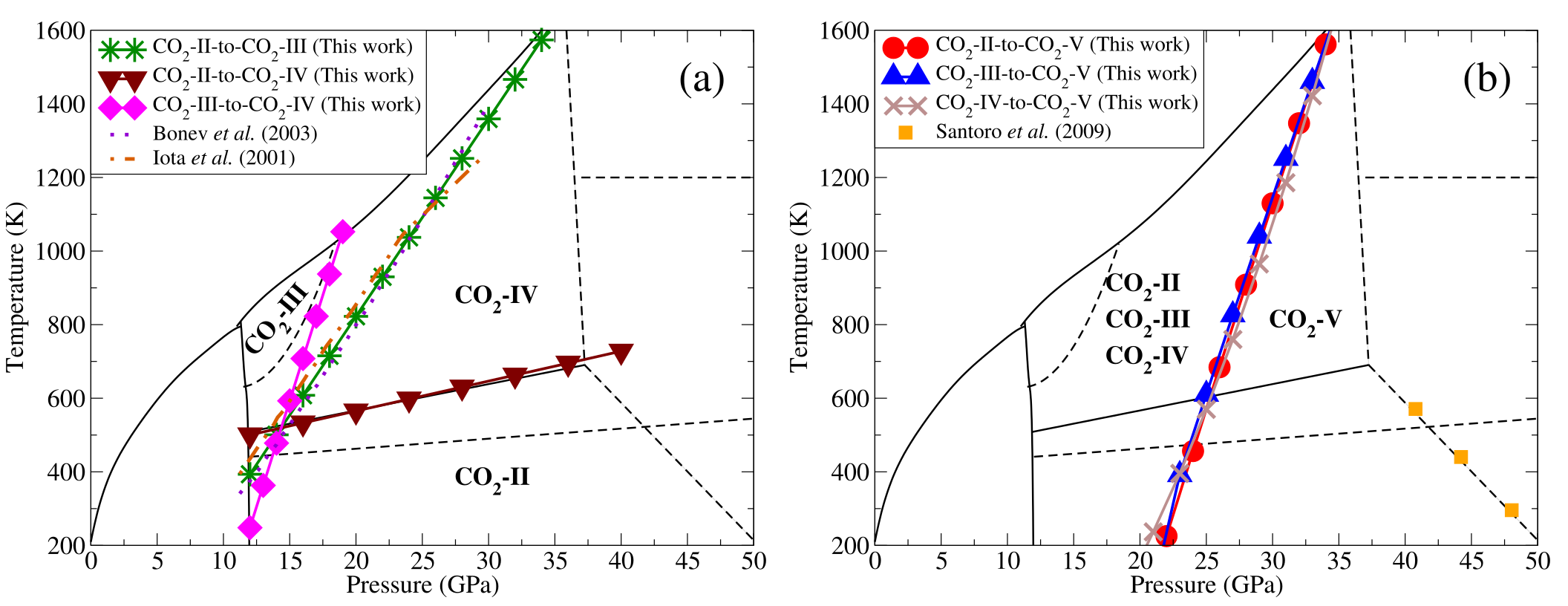}
\caption{(a) Phase boundaries between CO$_{2}$-II and CO$_{2}$-III (green, stars), CO$_{2}$-II and CO$_{2}$-IV (brown, down triangles), and CO$_{2}$-III and CO$_{2}$-IV (magenta, diamonds). Suggested boundaries reported by Iota \textit{et al.} \cite{Iota-2001} (experimental) and Bonev \textit{et al.}\cite{Bonev-2003} (theoretical), are shown in orange dot-dashed line and violet dotted line, respectively.
(b) Phase boundaries between molecular phases CO$_{2}$-II (red, circles), CO$_{2}$-III (blue, up triangles), and CO$_{2}$-IV (gray, crosses), and the non-molecular phase CO$_{2}$-V. Proposed limits of the kinetic region from experimental data from Ref. \cite{Santoro-2009} (orange squares) are also included.}
\label{fig:3}
\end{figure*}

\textit{Ab initio} electronic structure calculations were carried out using density functional theory (DFT) and the projector augmented wave (PAW) method, as implemented in the Quantum ESPRESSO suite \cite{Giannozzi-2009, Giannozzi-2017} with a kinetic energy cutoff of 200 Ry for the plane-wave basis set. The generalized gradient approximation (GGA) for the exchange-correlation energy was implemented using the Perdew-Burke-Ernzerhof functional \cite{Perdew-1996}. The Monkhorst-Pack method \cite{Pack-1976} was used to generate the $k$-points grids for sampling the Brillouin zone. Variable-cell optimization of all structural parameters was performed for the four phases in the range of pressures between 10 and 70 GPa. Density functional perturbation theory (DFPT) within the linear response scheme \cite{Baroni-2001} was used to calculate phonon frequencies at zero temperature. The zero-point energy and the finite-temperature contributions to the Helmholtz free energy were computed in the quasi-harmonic approximation (QHA) \cite{Leibfried-1961,Baroni-2010}. For the construction of the pressure-temperature phase diagram, the Helmholtz free energy at different temperatures was fitted to a $3^{rd}$ order Birch-Murnaghan equation of state. Finally, the Gibbs free energy was calculated as:
\begin{eqnarray}
G(P,T) &=& F[V(P,T),T] + PV(P,T),
\label{eq:2}
\end{eqnarray}

Room-temperature equations of state obtained with the above approximations are compared with experimental data for phases CO$_{2}$-II, CO$_{2}$-III, CO$_{2}$-IV, and CO$_{2}$-V in Fig. \ref{fig:2}. The agreement is quite good and confirms the validity of the approach. Phase boundaries constructed based on the calculated Gibbs free energies are shown in Fig. \ref{fig:3}.

\begin{figure}[t!]
\centering
\includegraphics[width=1.0\columnwidth]{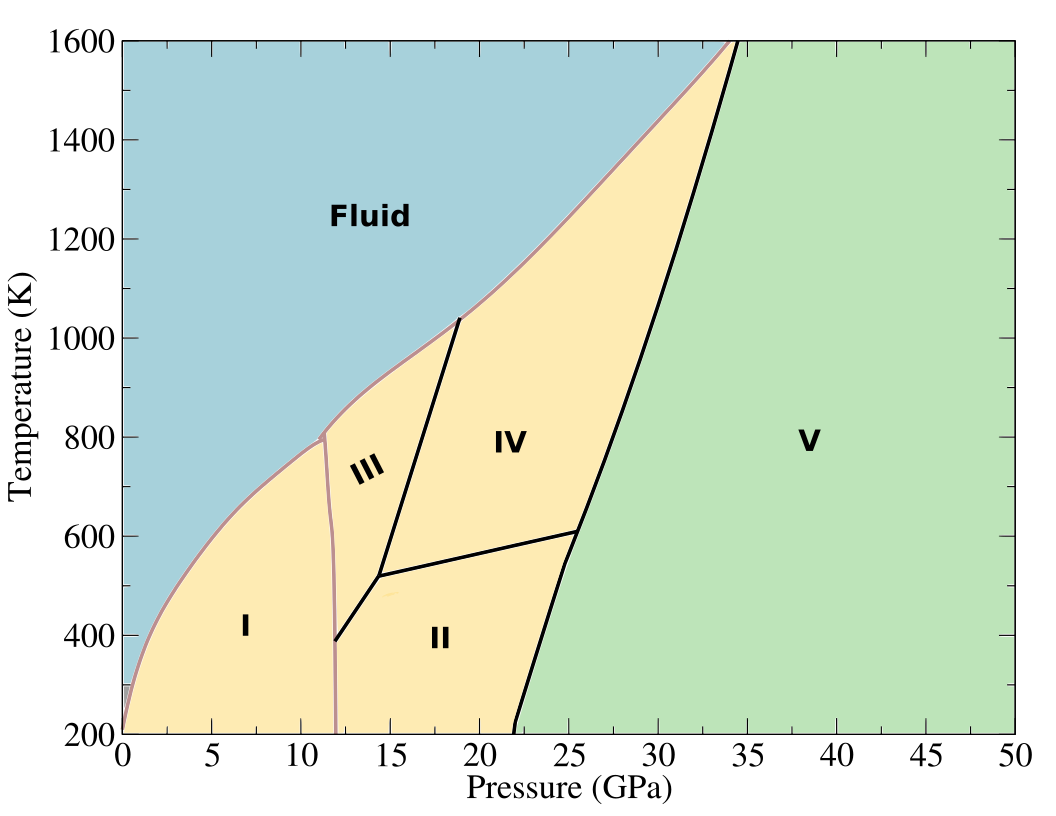}
\caption{Theoretical phase diagram for carbon dioxide at high pressure and temperature. Our calculated phase boundaries are shown with solid black lines, while previously reported thermodynamic boundaries are shown in gray. Yellow, green, and blue regions correspond to molecular, non-molecular, and fluid forms of CO$_{2}$.}
\label{fig:4}
\end{figure}

We begin our discussion with an analysis of the molecular solid  region of the phase diagram. This region is indicated in yellow in Fig. \ref{fig:1} and it contains the molecular phases I, II, III, and IV, and its upper bound in pressure coincides with the experimentally reported transitions to the non-molecular phases. Since phase I as well as its boundaries with the other phases are well known and constrained, we focus specifically on phases II, III, and IV, at pressures higher than 12 GPa. According to the enthalpy-pressure relations, with and without the zero-point energy contribution, at $T$ = 0 K CO$_{2}$-II is the most stable molecular phase in the pressure range considered, until the transition to CO$_{2}$-V. This indicates that the orthorhombic $Cmca$ structure (phase III) obtained experimentally from the compression of phase I is indeed only metastable at low temperatures. Notice that this remains true even after the inclusion of zero-point contributions, in agreement with previous reports \cite{Iota-2001,Bonev-2003}. Ref. \cite{Han-2019} proposes a calculated transition boundary between phases II and III in which CO$_{2}$-III is stable up to $\sim$ 570 K at 19 GPa, at odds with experimental observations where the kinetic transition from CO$_{2}$-III to CO$_{2}$-II occurs at much lower temperatures \cite{Iota-2001}. Instead, our calculations show that CO$_{2}$-III becomes more stable than CO$_{2}$-II at higher temperatures (solid green line with stars in Fig. \ref{fig:3}(a)). The transition temperature between CO$_{2}$-III and CO$_{2}$-II has a strong pressure dependence and reaches values in excess of 1000 K close to the boundary with the non-molecular phase V, with respect to its value close to CO$_{2}$-I. Comparing the free energies of CO$_{2}$-II and CO$_{2}$-IV we find that the boundary between phases II and IV (solid brown line with down triangles in Fig. \ref{fig:3}(a)) agrees quite well with experimental data \cite{Datchi-2009}. The weak pressure dependence of the II-IV boundary reduces the region of stability of phase II with respect to the starred green line in Fig. \ref{fig:3}(a), by confining it towards lower temperatures. Finally, we find that the boundary between CO$_{2}$-III and CO$_{2}$-IV (solid magenta line with diamonds in Fig. \ref{fig:3}(a)) is almost vertical, which restricts the domain of stability of phase III to a narrow window of pressure. We summarize the results of the free-energy calculations for the three molecular phases II, III, and IV in Fig. \ref{fig:4}. Phase I and II emerge as the only stable molecular phases of CO$_2$ from zero to ambient temperature. Phase III and IV are both stabilized by temperature, and phases II, III, and IV coexist at a triple point located at 15 GPa and 500 K.  

Our findings are also in agreement with simulations by Bonev \textit{et al.} \cite{Bonev-2003} which suggested that the $Cmca$ phase is a temperature stabilized form \cite{Bonev-2003}. Because the structure of phase IV was not known at the time, Bonev \textit{et al.} proposed a wider region of stability for CO$_{2}$-III. Interestingly, as can be seen in Fig. \ref{fig:3}(a), the $P-T$ region of stability of CO$_{2}$-III obtained from our calculations has a large overlap with the region of stability reported for the so-called phase VII of CO$_2$ \cite{Giordano-2007}. A recent theoretical work has shown that phase III and VII have, in fact, the same crystal structure (space group $Cmca$)\cite{Sontising-2017}. We therefore confirm that phase III is thermodynamically stable in the P-T region where phase VII has been reported. The observation of phase III outside this region (e.g. at ambient conditions, as a result of the compression of phase I) must be thus  attributed to kinetic effects. 
    
We now turn to the boundary between the molecular phases and non-molecular phase-V (Fig. \ref{fig:3}(b)). We find that at zero temperature the phase boundary between CO$_{2}$-II and CO$_{2}$-V is located at 21.5 GPa. The transition between (metastable) CO$_{2}$-III and CO$_{2}$-V would instead take place at 20.8 GPa in the absence of kinetic effects. This is in good agreement with previous theoretical works \cite{Oganov-2008,Gohr-2013,Yong-2016}. Phase boundaries between molecular phases and CO$_2$-V are rather insensitive to the choice of the molecular structure and they all have a positive slope, as already suggested \cite{Santoro-2004}. Considering that non-molecular phases are denser than molecular ones, a positive slope implies a decrease of entropy in going from molecular to non-molecular. This is not unexpected, given the stiffness of the non-molecular structure when compared with the molecular ones. Using the experimentally determined melting line, our calculations show a triple point between phases IV and V, and the liquid phase at 35 GPa and 1600 K. The calculations therefore suggest that molecular CO$_2$ could be stable up to pressures as high as 35 GPa, at high temperature. 

In summary, we have presented finite-temperature theoretical calculations in the quasi-harmonic approximation for various molecular and non-molecular solid forms of CO$_2$. The calculations aimed at resolving experimental uncertainties and inconsistencies due to kinetic effects and metastability. We find that the boundary between the molecular phases and phase V has a positive slope, and starts at 21.5 GPa at $T$ = 0 K. We also find that the phase diagram shows a triple point between phase IV, V, and the liquid phase at 35 GPa and 1600 K. This indicates that the non-molecular phase V has a broader region of stability than previously reported. We were able to reproduce the known thermodynamic boundary line between CO$_{2}$-II and CO$_{2}$-IV, confirming that kinetic effects are not relevant in that transition. Finally, it was shown that phase II is the most stable molecular phase at low temperatures, extending its region of stability to every $P-T$ condition where phase III has been reported experimentally. However, our results also show that CO$_{2}$-III is instead stabilized at high temperature and its stability region coincides with the $P-T$ conditions where phase VII has been reported experimentally, implying that phase III and phase VII are indeed the same.

We acknowledge the ISCRA program of CINECA for provision of computational time under proposal HP10CVCDOB. B.H.C-O thanks COLCIENCIAS for a graduate scholarship, also the support from the ICTP STEP programme is gratefully acknowledged. J.A.M. thanks the Vicerrector\'ia de Investigaciones of the Universidad de Cartagena, for the support of the Grupo de Modelado Computacional (COL0101016) through internal grants.

\bibliography{main.bib}

\begin{thebibliography}{38}%
\makeatletter
\providecommand \@ifxundefined [1]{%
 \@ifx{#1\undefined}
}%
\providecommand \@ifnum [1]{%
 \ifnum #1\expandafter \@firstoftwo
 \else \expandafter \@secondoftwo
 \fi
}%
\providecommand \@ifx [1]{%
 \ifx #1\expandafter \@firstoftwo
 \else \expandafter \@secondoftwo
 \fi
}%
\providecommand \natexlab [1]{#1}%
\providecommand \enquote  [1]{``#1''}%
\providecommand \bibnamefont  [1]{#1}%
\providecommand \bibfnamefont [1]{#1}%
\providecommand \citenamefont [1]{#1}%
\providecommand \href@noop [0]{\@secondoftwo}%
\providecommand \href [0]{\begingroup \@sanitize@url \@href}%
\providecommand \@href[1]{\@@startlink{#1}\@@href}%
\providecommand \@@href[1]{\endgroup#1\@@endlink}%
\providecommand \@sanitize@url [0]{\catcode `\\12\catcode `\$12\catcode
  `\&12\catcode `\#12\catcode `\^12\catcode `\_12\catcode `\%12\relax}%
\providecommand \@@startlink[1]{}%
\providecommand \@@endlink[0]{}%
\providecommand \url  [0]{\begingroup\@sanitize@url \@url }%
\providecommand \@url [1]{\endgroup\@href {#1}{\urlprefix }}%
\providecommand \urlprefix  [0]{URL }%
\providecommand \Eprint [0]{\href }%
\providecommand \doibase [0]{http://dx.doi.org/}%
\providecommand \selectlanguage [0]{\@gobble}%
\providecommand \bibinfo  [0]{\@secondoftwo}%
\providecommand \bibfield  [0]{\@secondoftwo}%
\providecommand \translation [1]{[#1]}%
\providecommand \BibitemOpen [0]{}%
\providecommand \bibitemStop [0]{}%
\providecommand \bibitemNoStop [0]{.\EOS\space}%
\providecommand \EOS [0]{\spacefactor3000\relax}%
\providecommand \BibitemShut  [1]{\csname bibitem#1\endcsname}%
\let\auto@bib@innerbib\@empty
\bibitem [{\citenamefont {Simon}\ and\ \citenamefont
  {Peters}(1980)}]{Simon-1980}%
  \BibitemOpen
  \bibfield  {author} {\bibinfo {author} {\bibfnamefont {A.}~\bibnamefont
  {Simon}}\ and\ \bibinfo {author} {\bibfnamefont {K.}~\bibnamefont {Peters}},\
  }\href@noop {} {\bibfield  {journal} {\bibinfo  {journal} {Acta
  Crystallographica Section B}\ }\textbf {\bibinfo {volume} {2}},\ \bibinfo
  {pages} {112750} (\bibinfo {year} {1980})}\BibitemShut {NoStop}%
\bibitem [{\citenamefont {Downs}\ and\ \citenamefont
  {Somayazulu}(1998)}]{Downs-1998}%
  \BibitemOpen
  \bibfield  {author} {\bibinfo {author} {\bibfnamefont {R.~T.}\ \bibnamefont
  {Downs}}\ and\ \bibinfo {author} {\bibfnamefont {M.~S.}\ \bibnamefont
  {Somayazulu}},\ }\href@noop {} {\bibfield  {journal} {\bibinfo  {journal}
  {Acta Crystallographica Section C}\ }\textbf {\bibinfo {volume} {54}},\
  \bibinfo {pages} {897} (\bibinfo {year} {1998})}\BibitemShut {NoStop}%
\bibitem [{\citenamefont {Aoki}\ \emph {et~al.}(1994)\citenamefont {Aoki},
  \citenamefont {Yamawaki}, \citenamefont {Sakashita}, \citenamefont {Gotoh},\
  and\ \citenamefont {Takemura}}]{Aoki-1994}%
  \BibitemOpen
  \bibfield  {author} {\bibinfo {author} {\bibfnamefont {K.}~\bibnamefont
  {Aoki}}, \bibinfo {author} {\bibfnamefont {H.}~\bibnamefont {Yamawaki}},
  \bibinfo {author} {\bibfnamefont {M.}~\bibnamefont {Sakashita}}, \bibinfo
  {author} {\bibfnamefont {Y.}~\bibnamefont {Gotoh}}, \ and\ \bibinfo {author}
  {\bibfnamefont {K.}~\bibnamefont {Takemura}},\ }\href@noop {} {\bibfield
  {journal} {\bibinfo  {journal} {Science}\ }\textbf {\bibinfo {volume}
  {263}},\ \bibinfo {pages} {356} (\bibinfo {year} {1994})}\BibitemShut
  {NoStop}%
\bibitem [{\citenamefont {Gimondi}\ and\ \citenamefont
  {Salvalaglio}(2017)}]{Gimondi-2017}%
  \BibitemOpen
  \bibfield  {author} {\bibinfo {author} {\bibfnamefont {I.}~\bibnamefont
  {Gimondi}}\ and\ \bibinfo {author} {\bibfnamefont {M.}~\bibnamefont
  {Salvalaglio}},\ }\href@noop {} {\bibfield  {journal} {\bibinfo  {journal}
  {The Journal of Chemical Physics}\ }\textbf {\bibinfo {volume} {147}},\
  \bibinfo {pages} {114502} (\bibinfo {year} {2017})}\BibitemShut {NoStop}%
\bibitem [{\citenamefont {Iota}\ and\ \citenamefont {Yoo}(2001)}]{Iota-2001}%
  \BibitemOpen
  \bibfield  {author} {\bibinfo {author} {\bibfnamefont {V.}~\bibnamefont
  {Iota}}\ and\ \bibinfo {author} {\bibfnamefont {C.~S.}\ \bibnamefont {Yoo}},\
  }\href@noop {} {\bibfield  {journal} {\bibinfo  {journal} {Physical Review
  Letters}\ }\textbf {\bibinfo {volume} {86}},\ \bibinfo {pages} {5922}
  (\bibinfo {year} {2001})}\BibitemShut {NoStop}%
\bibitem [{\citenamefont {Gorelli}\ \emph {et~al.}(2004)\citenamefont
  {Gorelli}, \citenamefont {Giordano}, \citenamefont {Salvi},\ and\
  \citenamefont {Bini}}]{Gorelli-2004}%
  \BibitemOpen
  \bibfield  {author} {\bibinfo {author} {\bibfnamefont {F.~A.}\ \bibnamefont
  {Gorelli}}, \bibinfo {author} {\bibfnamefont {V.~M.}\ \bibnamefont
  {Giordano}}, \bibinfo {author} {\bibfnamefont {P.~R.}\ \bibnamefont {Salvi}},
  \ and\ \bibinfo {author} {\bibfnamefont {R.}~\bibnamefont {Bini}},\
  }\href@noop {} {\bibfield  {journal} {\bibinfo  {journal} {Physical Review
  Letters}\ }\textbf {\bibinfo {volume} {93}},\ \bibinfo {pages} {3} (\bibinfo
  {year} {2004})}\BibitemShut {NoStop}%
\bibitem [{\citenamefont {Santoro}\ \emph {et~al.}(2006)\citenamefont
  {Santoro}, \citenamefont {Gorelli}, \citenamefont {Bini}, \citenamefont
  {Ruocco}, \citenamefont {Scandolo},\ and\ \citenamefont
  {Crichton}}]{Santoro-2006}%
  \BibitemOpen
  \bibfield  {author} {\bibinfo {author} {\bibfnamefont {M.}~\bibnamefont
  {Santoro}}, \bibinfo {author} {\bibfnamefont {F.~A.}\ \bibnamefont
  {Gorelli}}, \bibinfo {author} {\bibfnamefont {R.}~\bibnamefont {Bini}},
  \bibinfo {author} {\bibfnamefont {G.}~\bibnamefont {Ruocco}}, \bibinfo
  {author} {\bibfnamefont {S.}~\bibnamefont {Scandolo}}, \ and\ \bibinfo
  {author} {\bibfnamefont {W.~A.}\ \bibnamefont {Crichton}},\ }\href@noop {}
  {\bibfield  {journal} {\bibinfo  {journal} {Nature}\ }\textbf {\bibinfo
  {volume} {441}},\ \bibinfo {pages} {857} (\bibinfo {year}
  {2006})}\BibitemShut {NoStop}%
\bibitem [{\citenamefont {Han}\ \emph {et~al.}(2019)\citenamefont {Han},
  \citenamefont {Liu}, \citenamefont {Huang}, \citenamefont {He},\ and\
  \citenamefont {Li}}]{Han-2019}%
  \BibitemOpen
  \bibfield  {author} {\bibinfo {author} {\bibfnamefont {Y.}~\bibnamefont
  {Han}}, \bibinfo {author} {\bibfnamefont {J.}~\bibnamefont {Liu}}, \bibinfo
  {author} {\bibfnamefont {L.}~\bibnamefont {Huang}}, \bibinfo {author}
  {\bibfnamefont {X.}~\bibnamefont {He}}, \ and\ \bibinfo {author}
  {\bibfnamefont {J.}~\bibnamefont {Li}},\ }\href@noop {} {\bibfield  {journal}
  {\bibinfo  {journal} {npj Quantum Materials}\ }\textbf {\bibinfo {volume}
  {4}},\ \bibinfo {pages} {1} (\bibinfo {year} {2019})}\BibitemShut {NoStop}%
\bibitem [{\citenamefont {Yoo}\ \emph {et~al.}(2002)\citenamefont {Yoo},
  \citenamefont {Kohlmann}, \citenamefont {Cynn}, \citenamefont {Nicol},
  \citenamefont {Iota},\ and\ \citenamefont {LeBihan}}]{Yoo-2002}%
  \BibitemOpen
  \bibfield  {author} {\bibinfo {author} {\bibfnamefont {C.~S.}\ \bibnamefont
  {Yoo}}, \bibinfo {author} {\bibfnamefont {H.}~\bibnamefont {Kohlmann}},
  \bibinfo {author} {\bibfnamefont {H.}~\bibnamefont {Cynn}}, \bibinfo {author}
  {\bibfnamefont {M.~F.}\ \bibnamefont {Nicol}}, \bibinfo {author}
  {\bibfnamefont {V.}~\bibnamefont {Iota}}, \ and\ \bibinfo {author}
  {\bibfnamefont {T.}~\bibnamefont {LeBihan}},\ }\href@noop {} {\bibfield
  {journal} {\bibinfo  {journal} {Physical Review B}\ }\textbf {\bibinfo
  {volume} {65}},\ \bibinfo {pages} {104103} (\bibinfo {year}
  {2002})}\BibitemShut {NoStop}%
\bibitem [{\citenamefont {Datchi}\ \emph {et~al.}(2014)\citenamefont {Datchi},
  \citenamefont {Mallick}, \citenamefont {Salamat}, \citenamefont {Rousse},
  \citenamefont {Ninet}, \citenamefont {Garbarino}, \citenamefont {Bouvier},\
  and\ \citenamefont {Mezouar}}]{Datchi-2014}%
  \BibitemOpen
  \bibfield  {author} {\bibinfo {author} {\bibfnamefont {F.}~\bibnamefont
  {Datchi}}, \bibinfo {author} {\bibfnamefont {B.}~\bibnamefont {Mallick}},
  \bibinfo {author} {\bibfnamefont {A.}~\bibnamefont {Salamat}}, \bibinfo
  {author} {\bibfnamefont {G.}~\bibnamefont {Rousse}}, \bibinfo {author}
  {\bibfnamefont {S.}~\bibnamefont {Ninet}}, \bibinfo {author} {\bibfnamefont
  {G.}~\bibnamefont {Garbarino}}, \bibinfo {author} {\bibfnamefont
  {P.}~\bibnamefont {Bouvier}}, \ and\ \bibinfo {author} {\bibfnamefont
  {M.}~\bibnamefont {Mezouar}},\ }\href@noop {} {\bibfield  {journal} {\bibinfo
   {journal} {Physical Review B}\ }\textbf {\bibinfo {volume} {89}},\ \bibinfo
  {pages} {144101} (\bibinfo {year} {2014})}\BibitemShut {NoStop}%
\bibitem [{\citenamefont {Iota}\ \emph {et~al.}(2007)\citenamefont {Iota},
  \citenamefont {Yoo}, \citenamefont {Klepeis}, \citenamefont {Jenei},
  \citenamefont {Evans},\ and\ \citenamefont {Cynn}}]{Iota-2007}%
  \BibitemOpen
  \bibfield  {author} {\bibinfo {author} {\bibfnamefont {V.}~\bibnamefont
  {Iota}}, \bibinfo {author} {\bibfnamefont {C.~S.}\ \bibnamefont {Yoo}},
  \bibinfo {author} {\bibfnamefont {J.~H.}\ \bibnamefont {Klepeis}}, \bibinfo
  {author} {\bibfnamefont {Z.}~\bibnamefont {Jenei}}, \bibinfo {author}
  {\bibfnamefont {W.}~\bibnamefont {Evans}}, \ and\ \bibinfo {author}
  {\bibfnamefont {H.}~\bibnamefont {Cynn}},\ }\href@noop {} {\bibfield
  {journal} {\bibinfo  {journal} {Nature Materials}\ }\textbf {\bibinfo
  {volume} {6}},\ \bibinfo {pages} {34} (\bibinfo {year} {2007})}\BibitemShut
  {NoStop}%
\bibitem [{\citenamefont {Datchi}\ \emph {et~al.}(2009)\citenamefont {Datchi},
  \citenamefont {Giordano}, \citenamefont {Munsch},\ and\ \citenamefont
  {Saitta}}]{Datchi-2009}%
  \BibitemOpen
  \bibfield  {author} {\bibinfo {author} {\bibfnamefont {F.}~\bibnamefont
  {Datchi}}, \bibinfo {author} {\bibfnamefont {V.~M.}\ \bibnamefont
  {Giordano}}, \bibinfo {author} {\bibfnamefont {P.}~\bibnamefont {Munsch}}, \
  and\ \bibinfo {author} {\bibfnamefont {A.~M.}\ \bibnamefont {Saitta}},\
  }\href@noop {} {\bibfield  {journal} {\bibinfo  {journal} {Physical Review
  Letters}\ }\textbf {\bibinfo {volume} {103}},\ \bibinfo {pages} {1} (\bibinfo
  {year} {2009})}\BibitemShut {NoStop}%
\bibitem [{\citenamefont {Giordano}\ and\ \citenamefont
  {Datchi}(2007)}]{Giordano-2007}%
  \BibitemOpen
  \bibfield  {author} {\bibinfo {author} {\bibfnamefont {V.~M.}\ \bibnamefont
  {Giordano}}\ and\ \bibinfo {author} {\bibfnamefont {F.}~\bibnamefont
  {Datchi}},\ }\href@noop {} {\bibfield  {journal} {\bibinfo  {journal}
  {Europhysics Letters}\ }\textbf {\bibinfo {volume} {77}},\ \bibinfo {pages}
  {46002} (\bibinfo {year} {2007})}\BibitemShut {NoStop}%
\bibitem [{\citenamefont {Sontising}\ \emph {et~al.}(2017)\citenamefont
  {Sontising}, \citenamefont {Heit}, \citenamefont {McKinley},\ and\
  \citenamefont {Beran}}]{Sontising-2017}%
  \BibitemOpen
  \bibfield  {author} {\bibinfo {author} {\bibfnamefont {W.}~\bibnamefont
  {Sontising}}, \bibinfo {author} {\bibfnamefont {Y.~N.}\ \bibnamefont {Heit}},
  \bibinfo {author} {\bibfnamefont {J.~L.}\ \bibnamefont {McKinley}}, \ and\
  \bibinfo {author} {\bibfnamefont {G.~J.~O.}\ \bibnamefont {Beran}},\
  }\href@noop {} {\bibfield  {journal} {\bibinfo  {journal} {Chemical Science}\
  }\textbf {\bibinfo {volume} {8}},\ \bibinfo {pages} {7374} (\bibinfo {year}
  {2017})}\BibitemShut {NoStop}%
\bibitem [{\citenamefont {Iota}\ \emph {et~al.}(1999)\citenamefont {Iota},
  \citenamefont {Yoo},\ and\ \citenamefont {Cynn}}]{Iota-1999}%
  \BibitemOpen
  \bibfield  {author} {\bibinfo {author} {\bibfnamefont {V.}~\bibnamefont
  {Iota}}, \bibinfo {author} {\bibfnamefont {C.~S.}\ \bibnamefont {Yoo}}, \
  and\ \bibinfo {author} {\bibfnamefont {H.}~\bibnamefont {Cynn}},\ }\href@noop
  {} {\bibfield  {journal} {\bibinfo  {journal} {Science}\ }\textbf {\bibinfo
  {volume} {283}},\ \bibinfo {pages} {1510} (\bibinfo {year}
  {1999})}\BibitemShut {NoStop}%
\bibitem [{\citenamefont {Santoro}\ \emph {et~al.}(2012)\citenamefont
  {Santoro}, \citenamefont {Gorelli}, \citenamefont {Bini}, \citenamefont
  {Haines}, \citenamefont {Cambon}, \citenamefont {Levelut}, \citenamefont
  {Montoya},\ and\ \citenamefont {Scandolo}}]{Santoro-2012}%
  \BibitemOpen
  \bibfield  {author} {\bibinfo {author} {\bibfnamefont {M.}~\bibnamefont
  {Santoro}}, \bibinfo {author} {\bibfnamefont {F.~A.}\ \bibnamefont
  {Gorelli}}, \bibinfo {author} {\bibfnamefont {R.}~\bibnamefont {Bini}},
  \bibinfo {author} {\bibfnamefont {J.}~\bibnamefont {Haines}}, \bibinfo
  {author} {\bibfnamefont {O.}~\bibnamefont {Cambon}}, \bibinfo {author}
  {\bibfnamefont {C.}~\bibnamefont {Levelut}}, \bibinfo {author} {\bibfnamefont
  {J.~A.}\ \bibnamefont {Montoya}}, \ and\ \bibinfo {author} {\bibfnamefont
  {S.}~\bibnamefont {Scandolo}},\ }\href@noop {} {\bibfield  {journal}
  {\bibinfo  {journal} {Proceedings of the National Academy of Sciences}\
  }\textbf {\bibinfo {volume} {109}},\ \bibinfo {pages} {5176} (\bibinfo {year}
  {2012})}\BibitemShut {NoStop}%
\bibitem [{\citenamefont {Datchi}\ \emph {et~al.}(2012)\citenamefont {Datchi},
  \citenamefont {Mallick}, \citenamefont {Salamat},\ and\ \citenamefont
  {Ninet}}]{Datchi-2012}%
  \BibitemOpen
  \bibfield  {author} {\bibinfo {author} {\bibfnamefont {F.}~\bibnamefont
  {Datchi}}, \bibinfo {author} {\bibfnamefont {B.}~\bibnamefont {Mallick}},
  \bibinfo {author} {\bibfnamefont {A.}~\bibnamefont {Salamat}}, \ and\
  \bibinfo {author} {\bibfnamefont {S.}~\bibnamefont {Ninet}},\ }\href@noop {}
  {\bibfield  {journal} {\bibinfo  {journal} {Physical Review Letters}\
  }\textbf {\bibinfo {volume} {108}},\ \bibinfo {pages} {125701} (\bibinfo
  {year} {2012})}\BibitemShut {NoStop}%
\bibitem [{\citenamefont {Lee}\ \emph {et~al.}(2009)\citenamefont {Lee},
  \citenamefont {Montoya},\ and\ \citenamefont {Scandolo}}]{Lee-2009}%
  \BibitemOpen
  \bibfield  {author} {\bibinfo {author} {\bibfnamefont {M.~S.}\ \bibnamefont
  {Lee}}, \bibinfo {author} {\bibfnamefont {J.~A.}\ \bibnamefont {Montoya}}, \
  and\ \bibinfo {author} {\bibfnamefont {S.}~\bibnamefont {Scandolo}},\
  }\href@noop {} {\bibfield  {journal} {\bibinfo  {journal} {Physical Review
  B}\ }\textbf {\bibinfo {volume} {79}},\ \bibinfo {pages} {144102} (\bibinfo
  {year} {2009})}\BibitemShut {NoStop}%
\bibitem [{\citenamefont {Montoya}\ \emph {et~al.}(2008)\citenamefont
  {Montoya}, \citenamefont {Rousseau}, \citenamefont {Santoro}, \citenamefont
  {Gorelli},\ and\ \citenamefont {Scandolo}}]{Montoya-2008}%
  \BibitemOpen
  \bibfield  {author} {\bibinfo {author} {\bibfnamefont {J.~A.}\ \bibnamefont
  {Montoya}}, \bibinfo {author} {\bibfnamefont {R.}~\bibnamefont {Rousseau}},
  \bibinfo {author} {\bibfnamefont {M.}~\bibnamefont {Santoro}}, \bibinfo
  {author} {\bibfnamefont {F.}~\bibnamefont {Gorelli}}, \ and\ \bibinfo
  {author} {\bibfnamefont {S.}~\bibnamefont {Scandolo}},\ }\href@noop {}
  {\bibfield  {journal} {\bibinfo  {journal} {Phys. Rev. Lett.}\ }\textbf
  {\bibinfo {volume} {100}},\ \bibinfo {pages} {163002} (\bibinfo {year}
  {2008})}\BibitemShut {NoStop}%
\bibitem [{\citenamefont {Tschauner}\ \emph {et~al.}(2001)\citenamefont
  {Tschauner}, \citenamefont {Mao},\ and\ \citenamefont
  {Hemley}}]{Tschauner-2001}%
  \BibitemOpen
  \bibfield  {author} {\bibinfo {author} {\bibfnamefont {O.}~\bibnamefont
  {Tschauner}}, \bibinfo {author} {\bibfnamefont {H.~K.}\ \bibnamefont {Mao}},
  \ and\ \bibinfo {author} {\bibfnamefont {R.~J.}\ \bibnamefont {Hemley}},\
  }\href@noop {} {\bibfield  {journal} {\bibinfo  {journal} {Physical Review
  Letters}\ }\textbf {\bibinfo {volume} {87}},\ \bibinfo {pages} {075701}
  (\bibinfo {year} {2001})}\BibitemShut {NoStop}%
\bibitem [{\citenamefont {Litasov}\ \emph {et~al.}(2011)\citenamefont
  {Litasov}, \citenamefont {Goncharov},\ and\ \citenamefont
  {Hemley}}]{Litasov-2011}%
  \BibitemOpen
  \bibfield  {author} {\bibinfo {author} {\bibfnamefont {K.~D.}\ \bibnamefont
  {Litasov}}, \bibinfo {author} {\bibfnamefont {A.~F.}\ \bibnamefont
  {Goncharov}}, \ and\ \bibinfo {author} {\bibfnamefont {R.~J.}\ \bibnamefont
  {Hemley}},\ }\href@noop {} {\bibfield  {journal} {\bibinfo  {journal} {Earth
  and Planetary Science Letters}\ }\textbf {\bibinfo {volume} {309}},\ \bibinfo
  {pages} {318} (\bibinfo {year} {2011})}\BibitemShut {NoStop}%
\bibitem [{\citenamefont {Teweldeberhan}\ \emph {et~al.}(2013)\citenamefont
  {Teweldeberhan}, \citenamefont {Boates},\ and\ \citenamefont
  {Bonev}}]{Teweldeberhan-2013}%
  \BibitemOpen
  \bibfield  {author} {\bibinfo {author} {\bibfnamefont {A.}~\bibnamefont
  {Teweldeberhan}}, \bibinfo {author} {\bibfnamefont {B.}~\bibnamefont
  {Boates}}, \ and\ \bibinfo {author} {\bibfnamefont {S.}~\bibnamefont
  {Bonev}},\ }\href@noop {} {\bibfield  {journal} {\bibinfo  {journal} {Earth
  and Planetary Science Letters}\ }\textbf {\bibinfo {volume} {373}},\ \bibinfo
  {pages} {228} (\bibinfo {year} {2013})}\BibitemShut {NoStop}%
\bibitem [{\citenamefont {Dziubek}\ \emph {et~al.}(2018)\citenamefont
  {Dziubek}, \citenamefont {Scelta}, \citenamefont {Bini}, \citenamefont
  {Mezouar}, \citenamefont {Garbarino},\ and\ \citenamefont
  {Miletich}}]{Dziubek-2018}%
  \BibitemOpen
  \bibfield  {author} {\bibinfo {author} {\bibfnamefont {M.}~\bibnamefont
  {Dziubek}, \bibfnamefont {Kamil F.~andEnde}}, \bibinfo {author}
  {\bibfnamefont {D.}~\bibnamefont {Scelta}}, \bibinfo {author} {\bibfnamefont
  {R.}~\bibnamefont {Bini}}, \bibinfo {author} {\bibfnamefont {M.}~\bibnamefont
  {Mezouar}}, \bibinfo {author} {\bibfnamefont {G.}~\bibnamefont {Garbarino}},
  \ and\ \bibinfo {author} {\bibfnamefont {R.}~\bibnamefont {Miletich}},\
  }\href@noop {} {\bibfield  {journal} {\bibinfo  {journal} {Nature
  Communications}\ }\textbf {\bibinfo {volume} {9}},\ \bibinfo {pages} {3148}
  (\bibinfo {year} {2018})}\BibitemShut {NoStop}%
\bibitem [{\citenamefont {Seto}\ \emph {et~al.}(2010)\citenamefont {Seto},
  \citenamefont {Nishio-Hamane}, \citenamefont {Nagai}, \citenamefont {Sata},\
  and\ \citenamefont {Fujino}}]{Seto-2010}%
  \BibitemOpen
  \bibfield  {author} {\bibinfo {author} {\bibfnamefont {Y.}~\bibnamefont
  {Seto}}, \bibinfo {author} {\bibfnamefont {D.}~\bibnamefont {Nishio-Hamane}},
  \bibinfo {author} {\bibfnamefont {T.}~\bibnamefont {Nagai}}, \bibinfo
  {author} {\bibfnamefont {N.}~\bibnamefont {Sata}}, \ and\ \bibinfo {author}
  {\bibfnamefont {K.}~\bibnamefont {Fujino}},\ }\href@noop {} {\bibfield
  {journal} {\bibinfo  {journal} {Journal of Physics: Conference Series}\
  }\textbf {\bibinfo {volume} {215}},\ \bibinfo {pages} {012015} (\bibinfo
  {year} {2010})}\BibitemShut {NoStop}%
\bibitem [{\citenamefont {Datchi}\ \emph {et~al.}(2016)\citenamefont {Datchi},
  \citenamefont {Weck}, \citenamefont {Saitta}, \citenamefont {Raza},
  \citenamefont {Garbarino}, \citenamefont {Spaulding}, \citenamefont
  {Queyroux},\ and\ \citenamefont {Mezouar}}]{Datchi-2016}%
  \BibitemOpen
  \bibfield  {author} {\bibinfo {author} {\bibfnamefont {F.}~\bibnamefont
  {Datchi}}, \bibinfo {author} {\bibfnamefont {G.}~\bibnamefont {Weck}},
  \bibinfo {author} {\bibfnamefont {A.~M.}\ \bibnamefont {Saitta}}, \bibinfo
  {author} {\bibfnamefont {Z.}~\bibnamefont {Raza}}, \bibinfo {author}
  {\bibfnamefont {S.}~\bibnamefont {Garbarino}, \bibfnamefont {G.~Ninet}},
  \bibinfo {author} {\bibfnamefont {D.~K.}\ \bibnamefont {Spaulding}}, \bibinfo
  {author} {\bibfnamefont {J.~A.}\ \bibnamefont {Queyroux}}, \ and\ \bibinfo
  {author} {\bibfnamefont {M.}~\bibnamefont {Mezouar}},\ }\href@noop {}
  {\bibfield  {journal} {\bibinfo  {journal} {Physical Review B}\ }\textbf
  {\bibinfo {volume} {94}},\ \bibinfo {pages} {014201} (\bibinfo {year}
  {2016})}\BibitemShut {NoStop}%
\bibitem [{\citenamefont {Yong}\ \emph {et~al.}(2016)\citenamefont {Yong},
  \citenamefont {Liu}, \citenamefont {Wu}, \citenamefont {Yao}, \citenamefont
  {Tse}, \citenamefont {Dias},\ and\ \citenamefont {Yoo}}]{Yong-2016}%
  \BibitemOpen
  \bibfield  {author} {\bibinfo {author} {\bibfnamefont {X.}~\bibnamefont
  {Yong}}, \bibinfo {author} {\bibfnamefont {H.}~\bibnamefont {Liu}}, \bibinfo
  {author} {\bibfnamefont {M.}~\bibnamefont {Wu}}, \bibinfo {author}
  {\bibfnamefont {Y.}~\bibnamefont {Yao}}, \bibinfo {author} {\bibfnamefont
  {J.~S.}\ \bibnamefont {Tse}}, \bibinfo {author} {\bibfnamefont
  {R.}~\bibnamefont {Dias}}, \ and\ \bibinfo {author} {\bibfnamefont {C.~S.}\
  \bibnamefont {Yoo}},\ }\href@noop {} {\bibfield  {journal} {\bibinfo
  {journal} {Proceedings of the National Academy of Sciences}\ }\textbf
  {\bibinfo {volume} {113}},\ \bibinfo {pages} {11110} (\bibinfo {year}
  {2016})}\BibitemShut {NoStop}%
\bibitem [{\citenamefont {Santoro}\ \emph {et~al.}(2004)\citenamefont
  {Santoro}, \citenamefont {Lin}, \citenamefont {Mao},\ and\ \citenamefont
  {Hemley}}]{Santoro-2004}%
  \BibitemOpen
  \bibfield  {author} {\bibinfo {author} {\bibfnamefont {M.}~\bibnamefont
  {Santoro}}, \bibinfo {author} {\bibfnamefont {J.~F.}\ \bibnamefont {Lin}},
  \bibinfo {author} {\bibfnamefont {H.~K.}\ \bibnamefont {Mao}}, \ and\
  \bibinfo {author} {\bibfnamefont {R.~J.}\ \bibnamefont {Hemley}},\
  }\href@noop {} {\bibfield  {journal} {\bibinfo  {journal} {The Journal of
  Chemical Physics}\ }\textbf {\bibinfo {volume} {121}},\ \bibinfo {pages}
  {2780} (\bibinfo {year} {2004})}\BibitemShut {NoStop}%
\bibitem [{\citenamefont {Santoro}\ and\ \citenamefont
  {Gorelli}(2009)}]{Santoro-2009}%
  \BibitemOpen
  \bibfield  {author} {\bibinfo {author} {\bibfnamefont {M.}~\bibnamefont
  {Santoro}}\ and\ \bibinfo {author} {\bibfnamefont {F.~A.}\ \bibnamefont
  {Gorelli}},\ }\href@noop {} {\bibfield  {journal} {\bibinfo  {journal}
  {Physical Review B}\ }\textbf {\bibinfo {volume} {80}},\ \bibinfo {pages}
  {184109} (\bibinfo {year} {2009})}\BibitemShut {NoStop}%
\bibitem [{\citenamefont {Oganov}\ \emph {et~al.}(2008)\citenamefont {Oganov},
  \citenamefont {Ono}, \citenamefont {Ma}, \citenamefont {Glass},\ and\
  \citenamefont {Garcia}}]{Oganov-2008}%
  \BibitemOpen
  \bibfield  {author} {\bibinfo {author} {\bibfnamefont {A.~R.}\ \bibnamefont
  {Oganov}}, \bibinfo {author} {\bibfnamefont {S.}~\bibnamefont {Ono}},
  \bibinfo {author} {\bibfnamefont {Y.}~\bibnamefont {Ma}}, \bibinfo {author}
  {\bibfnamefont {C.~W.}\ \bibnamefont {Glass}}, \ and\ \bibinfo {author}
  {\bibfnamefont {A.}~\bibnamefont {Garcia}},\ }\href@noop {} {\bibfield
  {journal} {\bibinfo  {journal} {Earth and Planetary Science Letters}\
  }\textbf {\bibinfo {volume} {273}},\ \bibinfo {pages} {38} (\bibinfo {year}
  {2008})}\BibitemShut {NoStop}%
\bibitem [{\citenamefont {Gohr}\ \emph {et~al.}(2013)\citenamefont {Gohr},
  \citenamefont {Grimme}, \citenamefont {Söhnel}, \citenamefont {Paulus},\
  and\ \citenamefont {Schwerdtfeger}}]{Gohr-2013}%
  \BibitemOpen
  \bibfield  {author} {\bibinfo {author} {\bibfnamefont {S.}~\bibnamefont
  {Gohr}}, \bibinfo {author} {\bibfnamefont {S.}~\bibnamefont {Grimme}},
  \bibinfo {author} {\bibfnamefont {T.}~\bibnamefont {Söhnel}}, \bibinfo
  {author} {\bibfnamefont {B.}~\bibnamefont {Paulus}}, \ and\ \bibinfo {author}
  {\bibfnamefont {P.}~\bibnamefont {Schwerdtfeger}},\ }\href@noop {} {\bibfield
   {journal} {\bibinfo  {journal} {The Journal of Chemical Physics}\ }\textbf
  {\bibinfo {volume} {139}},\ \bibinfo {pages} {174501} (\bibinfo {year}
  {2013})}\BibitemShut {NoStop}%
\bibitem [{\citenamefont {Bonev}\ \emph {et~al.}(2003)\citenamefont {Bonev},
  \citenamefont {Gygi}, \citenamefont {Ogitsu},\ and\ \citenamefont
  {Galli}}]{Bonev-2003}%
  \BibitemOpen
  \bibfield  {author} {\bibinfo {author} {\bibfnamefont {S.~A.}\ \bibnamefont
  {Bonev}}, \bibinfo {author} {\bibfnamefont {F.}~\bibnamefont {Gygi}},
  \bibinfo {author} {\bibfnamefont {T.}~\bibnamefont {Ogitsu}}, \ and\ \bibinfo
  {author} {\bibfnamefont {G.}~\bibnamefont {Galli}},\ }\href@noop {}
  {\bibfield  {journal} {\bibinfo  {journal} {Physical Review Letters}\
  }\textbf {\bibinfo {volume} {91}},\ \bibinfo {pages} {065501} (\bibinfo
  {year} {2003})}\BibitemShut {NoStop}%
\bibitem [{\citenamefont {Giannozzi}\ \emph {et~al.}(2009)\citenamefont
  {Giannozzi}, \citenamefont {Baroni}, \citenamefont {Bonini}, \citenamefont
  {Calandra}, \citenamefont {Car}, \citenamefont {Cavazzoni}, \citenamefont
  {Ceresoli}, \citenamefont {Chiarotti}, \citenamefont {Cococcioni},
  \citenamefont {Dabo}, \citenamefont {{Dal Corso}}, \citenamefont {{De
  Gironcoli}}, \citenamefont {Fabris}, \citenamefont {Fratesi}, \citenamefont
  {Gebauer}, \citenamefont {Gerstmann}, \citenamefont {Gougoussis},
  \citenamefont {Kokalj}, \citenamefont {Lazzeri}, \citenamefont
  {Martin-Samos}, \citenamefont {Marzari}, \citenamefont {Mauri}, \citenamefont
  {Mazzarello}, \citenamefont {Paolini}, \citenamefont {Pasquarello},
  \citenamefont {Paulatto}, \citenamefont {Sbraccia}, \citenamefont {Scandolo},
  \citenamefont {Sclauzero}, \citenamefont {Seitsonen}, \citenamefont
  {Smogunov}, \citenamefont {Umari},\ and\ \citenamefont
  {Wentzcovitch}}]{Giannozzi-2009}%
  \BibitemOpen
  \bibfield  {author} {\bibinfo {author} {\bibfnamefont {P.}~\bibnamefont
  {Giannozzi}}, \bibinfo {author} {\bibfnamefont {S.}~\bibnamefont {Baroni}},
  \bibinfo {author} {\bibfnamefont {N.}~\bibnamefont {Bonini}}, \bibinfo
  {author} {\bibfnamefont {M.}~\bibnamefont {Calandra}}, \bibinfo {author}
  {\bibfnamefont {R.}~\bibnamefont {Car}}, \bibinfo {author} {\bibfnamefont
  {C.}~\bibnamefont {Cavazzoni}}, \bibinfo {author} {\bibfnamefont
  {D.}~\bibnamefont {Ceresoli}}, \bibinfo {author} {\bibfnamefont {G.~L.}\
  \bibnamefont {Chiarotti}}, \bibinfo {author} {\bibfnamefont {M.}~\bibnamefont
  {Cococcioni}}, \bibinfo {author} {\bibfnamefont {I.}~\bibnamefont {Dabo}},
  \bibinfo {author} {\bibfnamefont {A.}~\bibnamefont {{Dal Corso}}}, \bibinfo
  {author} {\bibfnamefont {S.}~\bibnamefont {{De Gironcoli}}}, \bibinfo
  {author} {\bibfnamefont {S.}~\bibnamefont {Fabris}}, \bibinfo {author}
  {\bibfnamefont {G.}~\bibnamefont {Fratesi}}, \bibinfo {author} {\bibfnamefont
  {R.}~\bibnamefont {Gebauer}}, \bibinfo {author} {\bibfnamefont
  {U.}~\bibnamefont {Gerstmann}}, \bibinfo {author} {\bibfnamefont
  {C.}~\bibnamefont {Gougoussis}}, \bibinfo {author} {\bibfnamefont
  {A.}~\bibnamefont {Kokalj}}, \bibinfo {author} {\bibfnamefont
  {M.}~\bibnamefont {Lazzeri}}, \bibinfo {author} {\bibfnamefont
  {L.}~\bibnamefont {Martin-Samos}}, \bibinfo {author} {\bibfnamefont
  {N.}~\bibnamefont {Marzari}}, \bibinfo {author} {\bibfnamefont
  {F.}~\bibnamefont {Mauri}}, \bibinfo {author} {\bibfnamefont
  {R.}~\bibnamefont {Mazzarello}}, \bibinfo {author} {\bibfnamefont
  {S.}~\bibnamefont {Paolini}}, \bibinfo {author} {\bibfnamefont
  {A.}~\bibnamefont {Pasquarello}}, \bibinfo {author} {\bibfnamefont
  {L.}~\bibnamefont {Paulatto}}, \bibinfo {author} {\bibfnamefont
  {C.}~\bibnamefont {Sbraccia}}, \bibinfo {author} {\bibfnamefont
  {S.}~\bibnamefont {Scandolo}}, \bibinfo {author} {\bibfnamefont
  {G.}~\bibnamefont {Sclauzero}}, \bibinfo {author} {\bibfnamefont {A.~P.}\
  \bibnamefont {Seitsonen}}, \bibinfo {author} {\bibfnamefont {A.}~\bibnamefont
  {Smogunov}}, \bibinfo {author} {\bibfnamefont {P.}~\bibnamefont {Umari}}, \
  and\ \bibinfo {author} {\bibfnamefont {R.~M.}\ \bibnamefont {Wentzcovitch}},\
  }\href@noop {} {\bibfield  {journal} {\bibinfo  {journal} {Journal of
  Physics: Condensed Matter}\ }\textbf {\bibinfo {volume} {21}},\ \bibinfo
  {pages} {21832390} (\bibinfo {year} {2009})}\BibitemShut {NoStop}%
\bibitem [{\citenamefont {Giannozzi}\ \emph {et~al.}(2017)\citenamefont
  {Giannozzi}, \citenamefont {Andreussi}, \citenamefont {Brumme}, \citenamefont
  {Bunau}, \citenamefont {Buongiorno~Nardelli}, \citenamefont {Calandra},
  \citenamefont {Car}, \citenamefont {Cavazzoni}, \citenamefont {Ceresoli},
  \citenamefont {Cococcioni}, \citenamefont {Colonna}, \citenamefont
  {Carnimeo}, \citenamefont {{Dal Corso}}, \citenamefont {{de Gironcoli}},
  \citenamefont {Delugas}, \citenamefont {DiStasio}, \citenamefont {Ferretti},
  \citenamefont {Floris}, \citenamefont {Fratesi}, \citenamefont {Fugallo},
  \citenamefont {Gebauer}, \citenamefont {Gerstmann}, \citenamefont {Giustino},
  \citenamefont {Gorni}, \citenamefont {Jia}, \citenamefont {Kawamura},
  \citenamefont {Ko}, \citenamefont {Kokalj}, \citenamefont
  {Kü{\c{c}}ükbenli}, \citenamefont {Lazzeri}, \citenamefont {Marsili},
  \citenamefont {Marzari}, \citenamefont {Mauri}, \citenamefont {L.},
  \citenamefont {Nguyen}, \citenamefont {Otero-de-la Roza}, \citenamefont
  {Paulatto}, \citenamefont {Ponc{\'{e}}}, \citenamefont {Rocca}, \citenamefont
  {Sabatini}, \citenamefont {Santra}, \citenamefont {Schlipf}, \citenamefont
  {Seitsonen}, \citenamefont {Smogunov}, \citenamefont {Timrov}, \citenamefont
  {Thonhauser}, \citenamefont {Umari}, \citenamefont {Vast}, \citenamefont
  {Wu},\ and\ \citenamefont {Baroni}}]{Giannozzi-2017}%
  \BibitemOpen
  \bibfield  {author} {\bibinfo {author} {\bibfnamefont {P.}~\bibnamefont
  {Giannozzi}}, \bibinfo {author} {\bibfnamefont {O.}~\bibnamefont
  {Andreussi}}, \bibinfo {author} {\bibfnamefont {T.}~\bibnamefont {Brumme}},
  \bibinfo {author} {\bibfnamefont {O.}~\bibnamefont {Bunau}}, \bibinfo
  {author} {\bibfnamefont {M.}~\bibnamefont {Buongiorno~Nardelli}}, \bibinfo
  {author} {\bibfnamefont {M.}~\bibnamefont {Calandra}}, \bibinfo {author}
  {\bibfnamefont {R.}~\bibnamefont {Car}}, \bibinfo {author} {\bibfnamefont
  {C.}~\bibnamefont {Cavazzoni}}, \bibinfo {author} {\bibfnamefont
  {D.}~\bibnamefont {Ceresoli}}, \bibinfo {author} {\bibfnamefont
  {M.}~\bibnamefont {Cococcioni}}, \bibinfo {author} {\bibfnamefont
  {N.}~\bibnamefont {Colonna}}, \bibinfo {author} {\bibfnamefont
  {I.}~\bibnamefont {Carnimeo}}, \bibinfo {author} {\bibfnamefont
  {A.}~\bibnamefont {{Dal Corso}}}, \bibinfo {author} {\bibfnamefont
  {S.}~\bibnamefont {{de Gironcoli}}}, \bibinfo {author} {\bibfnamefont
  {P.}~\bibnamefont {Delugas}}, \bibinfo {author} {\bibfnamefont {R.~A.}\
  \bibnamefont {DiStasio}}, \bibinfo {author} {\bibfnamefont {A.}~\bibnamefont
  {Ferretti}}, \bibinfo {author} {\bibfnamefont {A.}~\bibnamefont {Floris}},
  \bibinfo {author} {\bibfnamefont {G.}~\bibnamefont {Fratesi}}, \bibinfo
  {author} {\bibfnamefont {G.}~\bibnamefont {Fugallo}}, \bibinfo {author}
  {\bibfnamefont {R.}~\bibnamefont {Gebauer}}, \bibinfo {author} {\bibfnamefont
  {U.}~\bibnamefont {Gerstmann}}, \bibinfo {author} {\bibfnamefont
  {F.}~\bibnamefont {Giustino}}, \bibinfo {author} {\bibfnamefont
  {T.}~\bibnamefont {Gorni}}, \bibinfo {author} {\bibfnamefont
  {J.}~\bibnamefont {Jia}}, \bibinfo {author} {\bibfnamefont {M.}~\bibnamefont
  {Kawamura}}, \bibinfo {author} {\bibfnamefont {H.-Y.}\ \bibnamefont {Ko}},
  \bibinfo {author} {\bibfnamefont {A.}~\bibnamefont {Kokalj}}, \bibinfo
  {author} {\bibfnamefont {E.}~\bibnamefont {Kü{\c{c}}ükbenli}}, \bibinfo
  {author} {\bibfnamefont {M.}~\bibnamefont {Lazzeri}}, \bibinfo {author}
  {\bibfnamefont {M.}~\bibnamefont {Marsili}}, \bibinfo {author} {\bibfnamefont
  {N.}~\bibnamefont {Marzari}}, \bibinfo {author} {\bibfnamefont
  {F.}~\bibnamefont {Mauri}}, \bibinfo {author} {\bibfnamefont {N.~N.}\
  \bibnamefont {L.}}, \bibinfo {author} {\bibfnamefont {H.-V.}\ \bibnamefont
  {Nguyen}}, \bibinfo {author} {\bibfnamefont {A.}~\bibnamefont {Otero-de-la
  Roza}}, \bibinfo {author} {\bibfnamefont {L.}~\bibnamefont {Paulatto}},
  \bibinfo {author} {\bibfnamefont {S.}~\bibnamefont {Ponc{\'{e}}}}, \bibinfo
  {author} {\bibfnamefont {D.}~\bibnamefont {Rocca}}, \bibinfo {author}
  {\bibfnamefont {R.}~\bibnamefont {Sabatini}}, \bibinfo {author}
  {\bibfnamefont {B.}~\bibnamefont {Santra}}, \bibinfo {author} {\bibfnamefont
  {M.}~\bibnamefont {Schlipf}}, \bibinfo {author} {\bibfnamefont {A.~P.}\
  \bibnamefont {Seitsonen}}, \bibinfo {author} {\bibfnamefont {A.}~\bibnamefont
  {Smogunov}}, \bibinfo {author} {\bibfnamefont {I.}~\bibnamefont {Timrov}},
  \bibinfo {author} {\bibfnamefont {T.}~\bibnamefont {Thonhauser}}, \bibinfo
  {author} {\bibfnamefont {P.}~\bibnamefont {Umari}}, \bibinfo {author}
  {\bibfnamefont {N.}~\bibnamefont {Vast}}, \bibinfo {author} {\bibfnamefont
  {X.}~\bibnamefont {Wu}}, \ and\ \bibinfo {author} {\bibfnamefont
  {S.}~\bibnamefont {Baroni}},\ }\href@noop {} {\bibfield  {journal} {\bibinfo
  {journal} {Journal of Physics: Condensed Matter}\ }\textbf {\bibinfo {volume}
  {29}},\ \bibinfo {pages} {465901} (\bibinfo {year} {2017})}\BibitemShut
  {NoStop}%
\bibitem [{\citenamefont {Perdew}\ \emph {et~al.}(1996)\citenamefont {Perdew},
  \citenamefont {Burke},\ and\ \citenamefont {Ernzerhof}}]{Perdew-1996}%
  \BibitemOpen
  \bibfield  {author} {\bibinfo {author} {\bibfnamefont {J.~P.}\ \bibnamefont
  {Perdew}}, \bibinfo {author} {\bibfnamefont {K.}~\bibnamefont {Burke}}, \
  and\ \bibinfo {author} {\bibfnamefont {M.}~\bibnamefont {Ernzerhof}},\
  }\href@noop {} {\bibfield  {journal} {\bibinfo  {journal} {Physical Review
  Letters}\ }\textbf {\bibinfo {volume} {77}},\ \bibinfo {pages} {3865}
  (\bibinfo {year} {1996})}\BibitemShut {NoStop}%
\bibitem [{\citenamefont {Pack}\ and\ \citenamefont
  {Monkhorst}(1977)}]{Pack-1976}%
  \BibitemOpen
  \bibfield  {author} {\bibinfo {author} {\bibfnamefont {J.~D.}\ \bibnamefont
  {Pack}}\ and\ \bibinfo {author} {\bibfnamefont {H.~J.}\ \bibnamefont
  {Monkhorst}},\ }\href@noop {} {\bibfield  {journal} {\bibinfo  {journal}
  {Physical Review B}\ }\textbf {\bibinfo {volume} {16}},\ \bibinfo {pages}
  {1748} (\bibinfo {year} {1977})}\BibitemShut {NoStop}%
\bibitem [{\citenamefont {Baroni}\ \emph {et~al.}(2001)\citenamefont {Baroni},
  \citenamefont {{de Gironcoli}}, \citenamefont {{Dal Corso}},\ and\
  \citenamefont {Giannozzi}}]{Baroni-2001}%
  \BibitemOpen
  \bibfield  {author} {\bibinfo {author} {\bibfnamefont {S.}~\bibnamefont
  {Baroni}}, \bibinfo {author} {\bibfnamefont {S.}~\bibnamefont {{de
  Gironcoli}}}, \bibinfo {author} {\bibfnamefont {A.}~\bibnamefont {{Dal
  Corso}}}, \ and\ \bibinfo {author} {\bibfnamefont {P.}~\bibnamefont
  {Giannozzi}},\ }\href@noop {} {\bibfield  {journal} {\bibinfo  {journal}
  {Reviews of Modern Physics}\ }\textbf {\bibinfo {volume} {73}},\ \bibinfo
  {pages} {515} (\bibinfo {year} {2001})}\BibitemShut {NoStop}%
\bibitem [{\citenamefont {Leibfried}\ and\ \citenamefont
  {Ludwig}(1961)}]{Leibfried-1961}%
  \BibitemOpen
  \bibfield  {author} {\bibinfo {author} {\bibfnamefont {G.}~\bibnamefont
  {Leibfried}}\ and\ \bibinfo {author} {\bibfnamefont {W.}~\bibnamefont
  {Ludwig}},\ }\href@noop {} {\bibfield  {journal} {\bibinfo  {journal}
  {Reviews in Mineralogy and Geochemistry}\ }\textbf {\bibinfo {volume} {12}},\
  \bibinfo {pages} {275} (\bibinfo {year} {1961})}\BibitemShut {NoStop}%
\bibitem [{\citenamefont {Baroni}\ \emph {et~al.}(2010)\citenamefont {Baroni},
  \citenamefont {Giannozzi},\ and\ \citenamefont {Isaev}}]{Baroni-2010}%
  \BibitemOpen
  \bibfield  {author} {\bibinfo {author} {\bibfnamefont {S.}~\bibnamefont
  {Baroni}}, \bibinfo {author} {\bibfnamefont {P.}~\bibnamefont {Giannozzi}}, \
  and\ \bibinfo {author} {\bibfnamefont {E.}~\bibnamefont {Isaev}},\
  }\href@noop {} {\bibfield  {journal} {\bibinfo  {journal} {Solid State
  Physics - Advances in Research and Applications}\ }\textbf {\bibinfo {volume}
  {71}},\ \bibinfo {pages} {39} (\bibinfo {year} {2010})}\BibitemShut {NoStop}%
\end{thebibliography}%

\end{document}